\documentclass{jpp}

\usepackage{natbib}
\usepackage[T1]{fontenc}
\usepackage{bm}
\usepackage{color}
\usepackage{graphicx}
\usepackage[breaklinks=true]{hyperref}
\hypersetup{
  unicode=false,          
  pdftoolbar=true,        
  pdfmenubar=true,        
  pdffitwindow=false,     
  pdfstartview={FitH},    
  pdfsubject={Subject},   
  pdfcreator={Creator},   
  pdfproducer={Producer}, 
  pdfkeywords={keyword1} {key2} {key3}, 
  pdfnewwindow=true,      
  colorlinks=true,        
  linkcolor=blue,         
  citecolor=blue,         
  filecolor=blue,         
  urlcolor=blue           
}

\title{Enhancement of laser-driven ion acceleration in non-periodic nanostructured targets}

\author{I.~Thiele\aff{1}\corresp{\email{illia-thiele@web.de}}, J.~Ferri\aff{1}\corresp{\email{julien.ferri@chalmers.se}}, E.~Siminos\aff{2}, L.~Gremillet\aff{3}, E.~Smetanina\aff{2}, A.~Dmitriev\aff{2}, G.~Cantono\aff{4}, C.-G. Wahlstr{\"o}m\aff{4} \and T.~F\"ul\"op\aff{1}} 

\affiliation{\aff{1}Department of Physics, Chalmers University of Technology,
 SE-41296 G\"{o}teborg, Sweden
   \aff{2}Department of Physics, Gothenburg University, SE-41296 G\"{o}teborg, Sweden  \aff{3}CEA, DAM, DIF, F-91297 Arpajon, France
   \aff{4}Department of Physics, Lund University, SE-22100 Lund, Sweden}

\begin{document}

\maketitle

\begin{abstract}
  Using particle-in-cell simulations, we demonstrate an improvement of the target normal sheath acceleration (TNSA) of protons in non-periodically nanostructured targets with micron-scale thickness. Compared to standard flat foils, an increase in the proton cutoff energy by up to a factor of two is observed in foils coated with nanocones or perforated with nanoholes. The latter nano-perforated foils yield the highest enhancement, which we show to be robust over a broad range of foil thicknesses and hole diameters. The improvement of TNSA performance results from more efficient hot-electron generation, caused by a more complex laser-electron interaction geometry and increased effective interaction area and duration. We show that TNSA is optimized for a nanohole distribution of relatively low areal density and that is not required to be periodic, thus relaxing the manufacturing constraints.
\end{abstract}

\section{Introduction\label{sec:int}}
\indent

Laser-driven ion acceleration has become a well established technique to produce compact, high-energy ion beams, owing to the ultra-strong accelerating fields that can be achieved at the surfaces of solid targets~\citep{Daido12, Macchi13}. Such ion sources show great potential for a number of applications ranging from radiography \citep{Romagnani2005} to nuclear photonics~\citep{Habs2011} and  proton
therapy \citep{Bulanov2002}. However, even though  proton  energies close to 100 MeV have been demonstrated in recent experiments using petawatt-class laser facilities~\citep{Wagner16, Higginson2018},  the few tens of MeV energies that are routinely attained using multi-terawatt-class laser systems are insufficient for many of the foreseen applications, which therefore limit the applicability of laser-driven ion sources. This spurs the development of novel schemes yielding significantly increased proton energies.

The most robust, and extensively investigated, acceleration scheme is the so-called  target-normal-sheath acceleration~(TNSA)~\citep{Snavely2000, Wilks01}, whereby surface ions are driven outwards by the charge separation field set up by the laser-accelerated relativistic electrons escaping into vacuum. 
Because of their large charge-to-mass ratio, the protons that are naturally present due to hydrogen-containing contaminants at the target surfaces respond the fastest to the electric sheath field, and reach the highest velocities. Their final energy spectrum has typically the form of a decreasing  exponential with a sharp high-energy cutoff. 

Different strategies have been explored in recent years to increase the proton cutoff energies resulting from TNSA. With micrometric foil targets, this requires enhancing the fast electron generation at the target front side. One option is to manipulate the laser temporal profile so as to create a preplasma with an optimal scale length~\citep{kaluza04,Nuter2008}, or to induce an optimal electromagnetic interference pattern~\citep{Ferri2019}. Another option is to modify the target properties: reduction of the target thickness~\citep{Neely06,Ogura12} or transverse size~\citep{Buffechoux10} thus results in higher proton energies and numbers.   An alternative, which is addressed in the present paper, is to employ nano- (or micro-)structured targets. While twofold increase in proton energy has been reported from foils with periodic surface structures~\citep{Margarone12,Ceccotti13}, little attention has been paid so far to the potential of non-periodic structures~\citep{Zigler13,Fedeli2018}. 
However, relaxing the constraint on the periodicity would enable simpler and more robust target fabrication methods~\citep{Langhammer07,Zigler11}, as is required to bring laser-driven proton sources closer to applications.

In this paper, we investigate by means of particle-in-cell~(PIC)
simulations, the potential of non-periodically structured targets to
enhance TNSA. Two target types are considered, consisting of a flat
foil either coated on the front surface with randomly positioned
nanocones (``nanocone targets'') or perforated by nanoholes
(``nanohole targets'').  In both cases the proton cutoff energy is
increased by up to a factor two compared with the case of flat foils.
The improvement is a result of to a more efficient hot-electron
generation mainly due to the increased effective laser-matter interaction
area and duration. We find that nanohole targets with relatively low
areal density give the highest enhancement. In  Sec.~\ref{sec:setup},
we describe the physical and numerical setup and in
Sec.~\ref{sec:orig} we compare the results obtained with the
structured targets to the flat-foil case and investigate the origin of
proton energy enhancement. In Sec.~\ref{sec:scan}, a parametric study
of the nanohole targets is presented. Finally, in Sec.~\ref{sec:sum}
we summarize our results.

\section{\label{sec:setup}Physical and numerical setups}
\indent

\begin{figure}
	\centering
	\includegraphics[width=0.99\columnwidth]{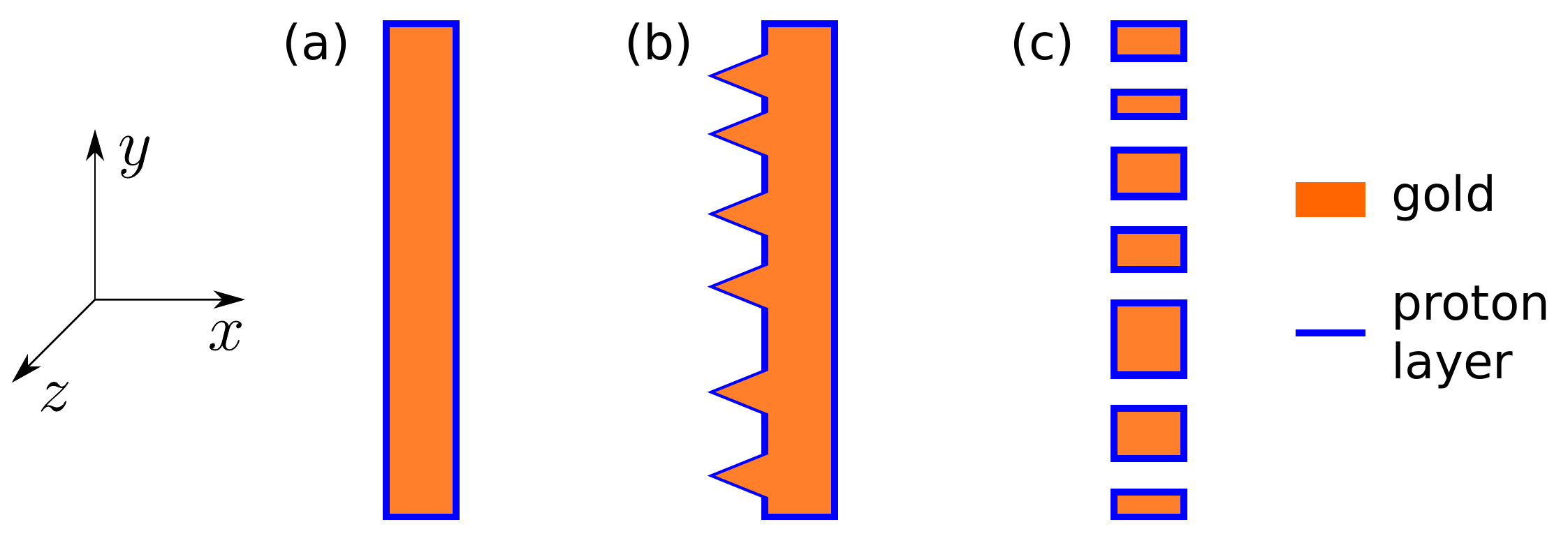}
	\caption{Schematic representation of the flat foil~(a), nanocone~(b) and nanohole~(c) targets. The targets are irradiated by the laser pulse from the left under an incidence angle of $45^\circ$. }
	\label{fig:spectra_cones_vs_holes1}
\end{figure}

We will investigate how the aforementioned two types of structured foils behave with
respect to proton acceleration by means of 2D PIC simulations with the
 {\sc smilei} code~\citep{SMILEI}. The considered structures are visualized in
Figs.~\ref{fig:spectra_cones_vs_holes1}(a-c).  The reference target is a
flat foil of thickness $d = 100\; \rm nm$. It is composed of gold atoms,
assumed to be 11-times ionized, with an ion number density of
$5.85\times 10^{22}\,\mathrm{cm}^{-3}$. Simulations with different 
ionization states of gold atoms (up to 29-times ionized) 
were also performed but did not result in significant differences 
in proton spectra. 
A $10\;\rm nm$ thin
proton-electron plasma layer with a number density of
$1.74\times 10^{23}\,\mathrm{cm}^{-3}$ is added on the foil surfaces to
model the hydrogen contaminants.

The nanocone target, sketched in
Fig.~\ref{fig:spectra_cones_vs_holes1}(b), is composed of the above-mentioned
flat foil coated with a distribution of cones, each having an opening
angle of $44^\circ$ and a variable base size of $b$ (specific values will be set below). The nanohole
target, displayed in Fig.~\ref{fig:spectra_cones_vs_holes1}(c),
consists of the reference foil pierced by holes of the same width and
location as the above nanocones.  The surfaces of both types of
structured targets are coated with a  proton layer. Such geometries are feasible to realize experimentally, for example, with colloidal lithography~\citep{Fredriksson07, Lodewijks16}.

 For the initialization of the
nanocones' positions, we choose the following model: Given the position $y_1$ of the first nanocone, the average distance $D = 2.24 b$ and the distance spread
parameter $s = D/2$, the position of the $i$th nanocone is set iteratively to $y_i=y_{i-1}+D+(u-1/2)s$ as long as
$y_i<y_{\rm max}$, where $y_{\rm max}$ is the maximal position and $u$ is a uniformly distributed random number taking
values from 0 to 1. This leads to an average cone coverage density of
$\rho=b/D=44\%$. The targets are located at $x=20\;\rm \mu m$.

The $p$-polarized laser pulse has a wavelength of $\lambda=0.8\,\mu$m
and a maximum intensity of $5\times 10^{19}$~W/cm$^2$, corresponding to
normalized vector potential $a_0=4.8$ and a peak electric field of
$194$~GV/cm. It has Gaussian space and time profiles with a focal
spot of $5\,\mu$m FWHM and a duration of 38~fs FWHM. It is incident on
the target from the lower left-hand side, at $45^\circ$ from the
surface normal. The peak intensity of the laser pulse reaches the foil after 175~fs in the simulation. In the following, we will set this as the time reference $t=0$.

The numerical discretization of the simulations is
$\delta x = \delta y = 5$~nm and $\delta t = 11.7$~as. We use 100
macro-particles per cell and per species in the bulk plasma, while the
surface proton-electron layers are represented by 1000 macro-particles
per cell and per species.

\section{\label{sec:orig}Enhancement of electron heating and ion acceleration}
\indent

Periodic cone  structures have been shown to enhance  proton
acceleration due to a modification of electron
trajectories, hence maximizing  laser absorption. Such enhancement has already been  discussed for periodic nanohole targets~\citep{Nodera08}, periodic nanobrush targets~\citep{yu12} and grating surfaces~\citep{Sgattoni2015,Blanco17}. Thus, it is interesting to investigate whether relaxing the restriction of periodicity would impact the acceleration process. Here, we consider a
non-periodic arrangement of the structures. The cone and hole base size in this
section is fixed to be $b=300$~nm.

\begin{figure}
	\centering
	\includegraphics[width=0.99\columnwidth]{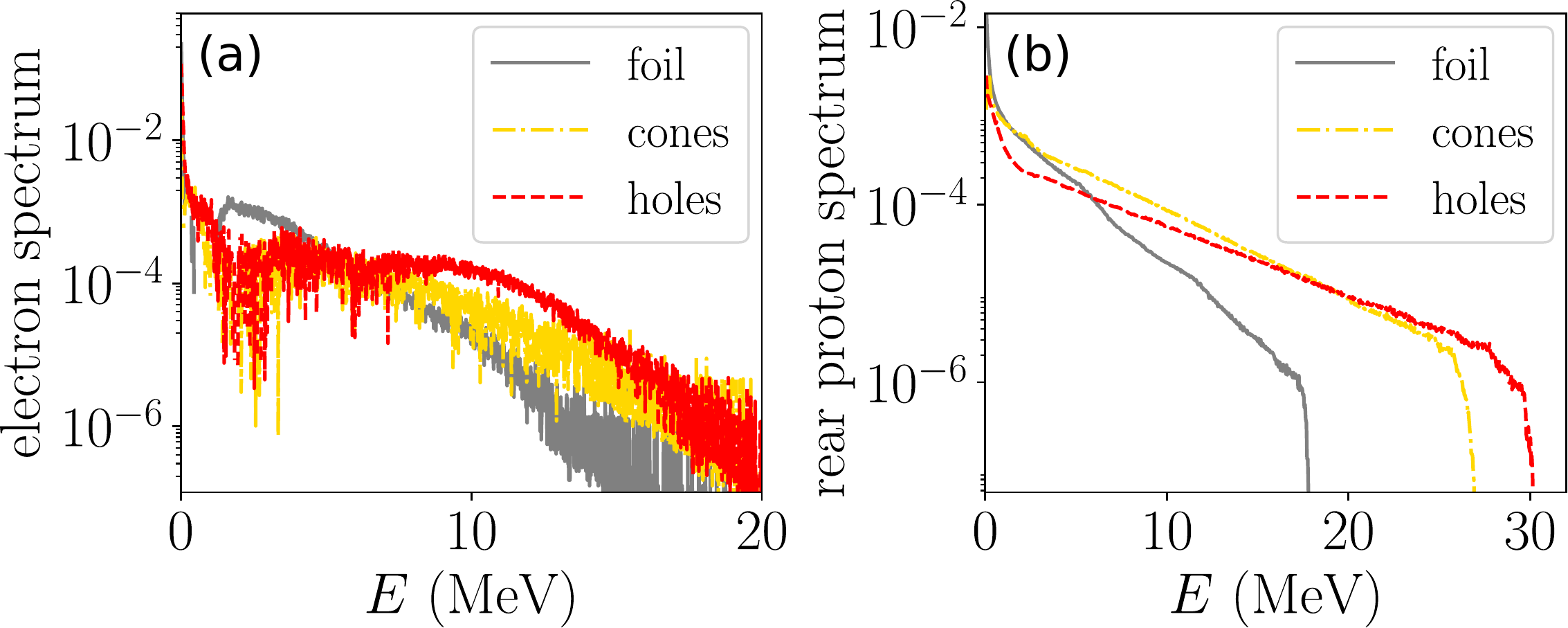}
	\caption{Energy spectra of electrons~(a) and rear-side protons~(b) from a flat foil (solid black), a nanocone (dashed-dotted yellow) and a nanohole (dashed red) targets of $d=100\,\rm nm$ thickness. In (a), only electrons located in the vacuum behind the target backside are considered. Electron spectra are recorded at $t=175$~fs and proton spectra at $t=455$~fs.}
	\label{fig:spectra_cones_vs_holes2}
\end{figure}

The strength of the rear-side sheath field that underpins TNSA, and which therefore determines the efficiency of the latter, is controlled by the energy density of the laser-generated hot electrons~\citep{Daido12, Macchi13}. 
Figure~\ref{fig:spectra_cones_vs_holes2}(a) plots the energy spectrum of the electrons located in the vacuum region behind the rear side of the target, recorded at $t=175$~fs. Those electrons mainly account for the generation of the sheath electric field in the early stages of TNSA, when the approximation of the 1D plasma expansion holds.
 Compared with the flat foil, both structured targets lead to a significantly increased amount of relativistic electrons above $7\,\rm MeV$, with the nanohole target yielding the largest enhancement -- by about an order of magnitude. 
\begin{figure}
	\centering
	\includegraphics[width=0.99\columnwidth]{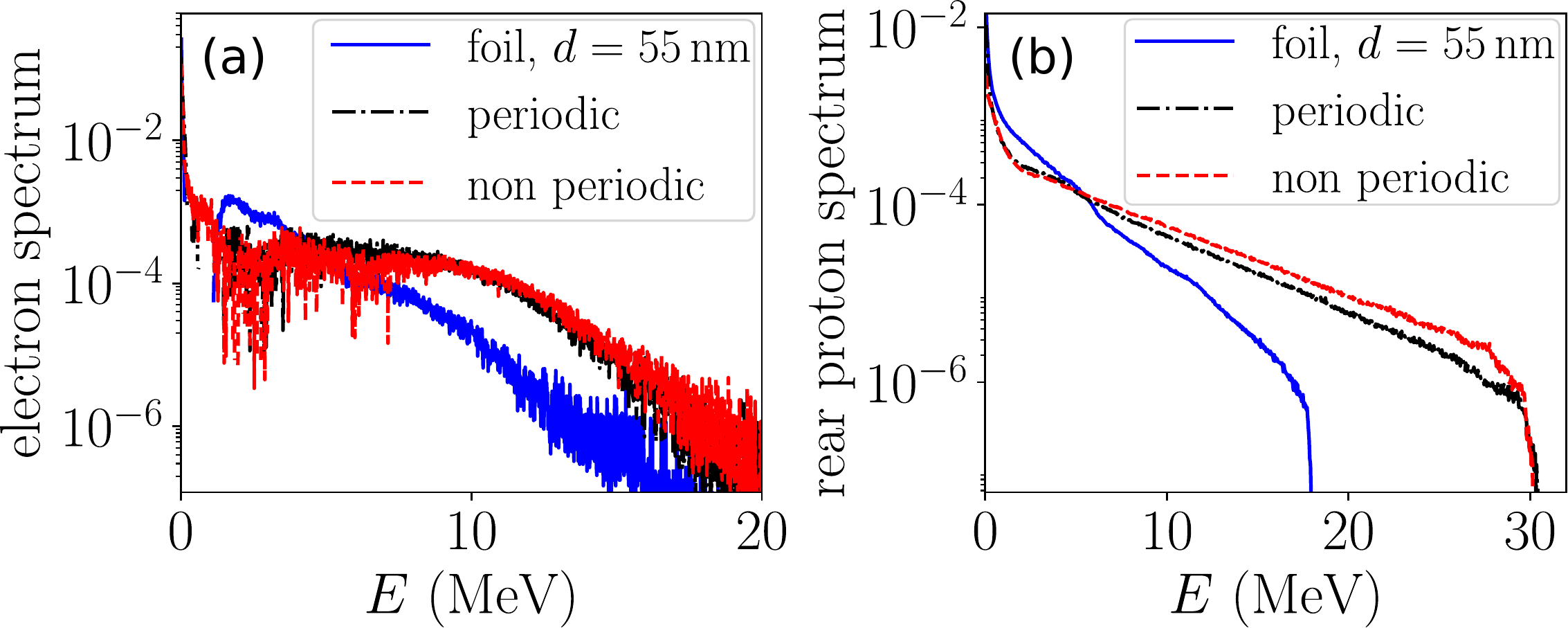}
	\caption{Energy spectra of the electrons~(a) and rear-protons~(b) from a flat foil (solid blue), a $d=100\,\rm nm$ periodic nanohole target (dashed-dotted black) and a $d=100\,\rm nm$ non-periodic nanohole target (dashed red). In (a), only electrons located in the vacuum behind the target backside are considered. Electron spectra are recorded at $t=175$~fs and proton spectra at $t=455$~fs.}
	\label{fig:spectra_periodic}
\end{figure}
\begin{figure*}
	\centering
	\includegraphics[width=0.99\columnwidth]{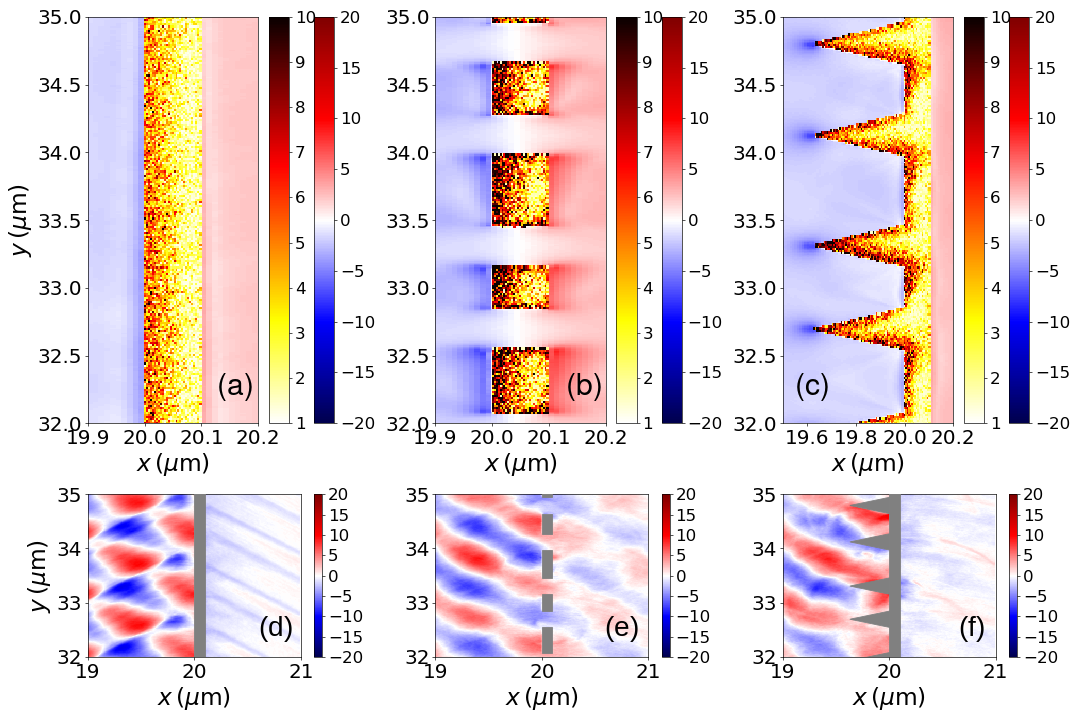}
	\caption{(a)-(c) Maximum $\gamma$ factor (locally averaged over the particle distribution) reached by electrons as a function of their initial position (yellow--red) and longitudinal electric field $E_x$ (units $m_ec\omega_0/e$, with $c$ the speed of light, $\omega_0$ the laser frequency, $m_e$ the electron mass and $e$ the elementary charge) at $t=0$, averaged over a laser period $T_0$ (blue--red). (d)-(f) Magnetic field $B_z$ (units $m_e\omega_0/e$) at $t=0$. The initial target densities are indicated in light gray, and correspond to $100\,\rm nm$-thick flat (a,d) nanohole (b,e) and nanocone targets (c,f).}
	\label{fig:initpos}
\end{figure*}

A similar behavior is found for the rear-side proton energy spectra, as plotted in Fig.~\ref{fig:spectra_cones_vs_holes2}(b) at $t=455~$fs. Both nanocone and nanohole targets give rise to much enhanced TNSA: the best performance is observed using nanoholes, with an almost doubled proton cutoff energy compared to that from the flat foil ($\sim 30\,\rm MeV$ protons vs $\sim 17\,\rm MeV$).

 Figure~\ref{fig:spectra_periodic} demonstrates that the electron and proton spectra are almost the same for periodic and non-periodic targets. It is thus to be expected that the individual structuring units, rather than their periodic arrangement (possibly leading to the excitation of surface plasma waves), are responsible for the enhancement. This is a favorable result from an experimental perspective since it reduces the target manufacturing constraints.

Moreover, Figs.~\ref{fig:spectra_periodic}(a,b) present the electron and rear-proton energy spectra obtained for a flat foil (blue solid line) of reduced thickness ($d=55\,\mathrm{nm}$) such that it contains the same total amount of matter as the $d=100\,\rm nm$ nanohole targets, whether periodic or not. This thinner flat foil produces particle energy spectra very similar to the $d=100\,\rm nm$ foil; hence, the enhanced performance of the nanohole targets cannot be ascribed to the direct effect of their reduced volume (or area in 2D) on the electron kinetic energy density (which would naturally increase assuming the same amount of laser energy is converted into hot electrons), but rather results from a strongly modified hot-electron dynamics. 

\begin{figure*}
	\centering
	\includegraphics[width=0.95\columnwidth]{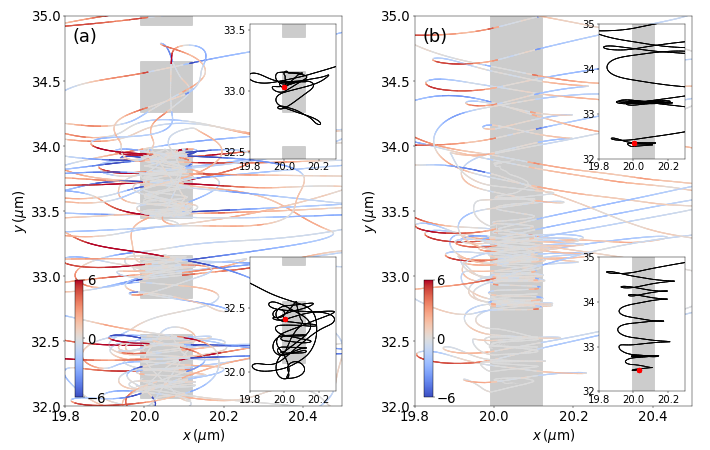}
	\caption{Trajectories of a few energetic electrons in the nanohole target (a) and the flat target (b). These electrons are selected randomly amongst those verifying $x>48~\mu$m at $t=175$~fs, and as long as they have reached a threshold energy $E_{th}$ during the simulation, with $E_{th} = 12$~MeV in the nanohole case and $E_{th} = 5$~MeV in the flat target case. The initial target density is indicated in light gray. The color of the trajectories (blue--red) represents the rate of change ($(p_xE_x+p_yE_y)/\gamma$) of electron energy (units $m_e^2c^2\omega_0/e$). Two trajectories of single electrons (black) are represented for each case in the different insets. The red dots indicate the initial positions of these particles.}
	\label{fig:track_part}
\end{figure*}

In order to gain insight into the electron energization process in the nanostructured target, we record the maximum Lorentz factor $\gamma$ reached by each electron during the simulation, and plot its locally averaged value as a function of the initial electron position [Fig.~\ref{fig:initpos}(a)-(c)]. In the case of the flat foil [Fig.~\ref{fig:initpos}(a)], the resulting map shows that, as expected, most of the accelerated electrons originate from a $\sim 25\,\rm nm$-thick surface layer at the directly irradiated front-side of the target (with mean energies $\sim 6\, \rm MeV$ being reached), and that rear-side electrons undergo negligible acceleration. 
In the case of the nanocone [Fig.~\ref{fig:initpos}(b)] and nanohole [Fig.~\ref{fig:initpos}(c)] targets, part of the highest energy electrons is stemming from additional regions, namely the nanohole walls and the nanocone sides. The nanostructuring of the target surface therefore increases the effective interaction area, leading to a larger number of hot electrons. The mean energy ($\sim 10\, \rm MeV$) reached by these electrons is also significantly larger than in the flat foil. 
These effects are supported by the local enhancement of the electrostatic field which appears when using nanostructured targets. The electrostatic field is indeed strongly enhanced at the corners of the nanoholes and at the tips of the nanocones, which correspond to the surfaces from which the most energetic electrons arise. Note that the value of the electrostatic field in these regions becomes of the same order as the laser field [Fig.~\ref{fig:initpos}(d)-(f)], which is also favorable for electron acceleration~\citep{Paradkar12}. Interestingly, when the target is nanostructured, the change of the laser field pattern on the front side of the target indicates a decrease in the laser reflection compared with the flat foil. In the case of the nanohole target, laser fields are propagating in the nanoholes, leading to partial transmission. 

This general behavior can be complemented by examining individual electron trajectories. We focus on those high-energy electrons breaking through the target rear side, and therefore contributing to the accelerating sheath field. 
 For this reason, Fig.~\ref{fig:track_part} plots the trajectories of a sample of the most energetic electrons in both the flat and nanohole targets -- with a lower energy cutoff for the selection of 12~MeV in the nanohole target [Fig.~\ref{fig:track_part}(a)] and of 5~MeV in the flat target case [Fig.~\ref{fig:track_part}(b)]. Only those electrons that are located at a longitudinal position $x\ge 48~\mu$m at $t=175$~fs are selected. The color of each trajectory is indexed on the rate of change of electron energy in the local electromagnetic fields. It can be seen that, on average, higher values are attained in the nanohole target. While the selected electrons mainly exhibit acceleration in the front- and rear-side vacuum regions, sizable energy transfer is also seen within the cavities, which is consistent with the laser being partially transmitted through the nanoholes [see Fig.~\ref{fig:initpos}(e)].

\begin{figure}
	\centering
	\includegraphics[width=0.6\columnwidth]{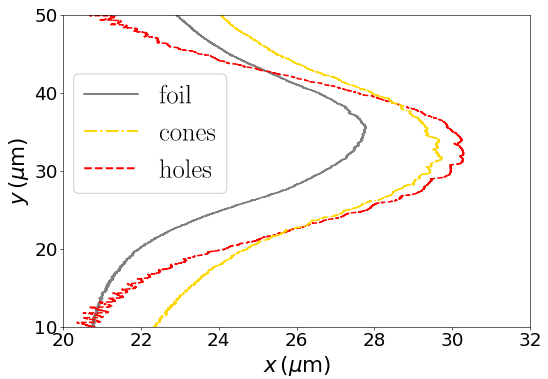}
	\caption{Position of the front of the accelerating rear-proton layer at $t=175$~fs for a flat foil (solid black), a nanocone (dashed-dotted yellow) and a nanohole (dashed red) targets of $d=100\,\rm nm$ thickness.}
	\label{fig:sheath}
\end{figure}

The sub-micron dense regions that make up the nanohole targets effectively behave as mass-limited targets~\citep{Psikal08,Buffechoux10}, which lead to efficient electrostatic confinement of the hot electrons. This is evidenced by the single particle trajectories displayed in the insets of Fig.~\ref{fig:track_part}: the nanohole target causes the electron to recirculate in both longitudinal (across the front and rear sides of the target) and transverse (across the nanohole walls) directions. A favorable consequence of this is a longer effective laser-electron interaction time. Also, the laser-electron interaction occurs under various geometrical conditions, and so with increased degrees of freedom. This results in a more complete exploration of the phase space, which ultimately allows the electrons to be accelerated to higher energies. Finally, being prevented from leaving the laser--irradiated region, the hot electrons are able to sustain a strong sheath field over longer times. The confinement of the hot electrons leads to a reduced transverse extent of the sheath field, and therefore of the expanding proton cloud. This can be seen in Figure~\ref{fig:sheath}, which shows the boundary of the proton cloud as resulting from the three target types: the nanohole target gives rise to a more narrow proton distribution. Note that this also corresponds to a more divergent proton beam; the divergence is multiplied by a factor $\sim 2$ between the flat and the nanohole targets.

\section{\label{sec:scan}Parametric scan for the nanohole targets}
\indent

\begin{figure*}
	\centering
	\includegraphics[width=1\columnwidth]{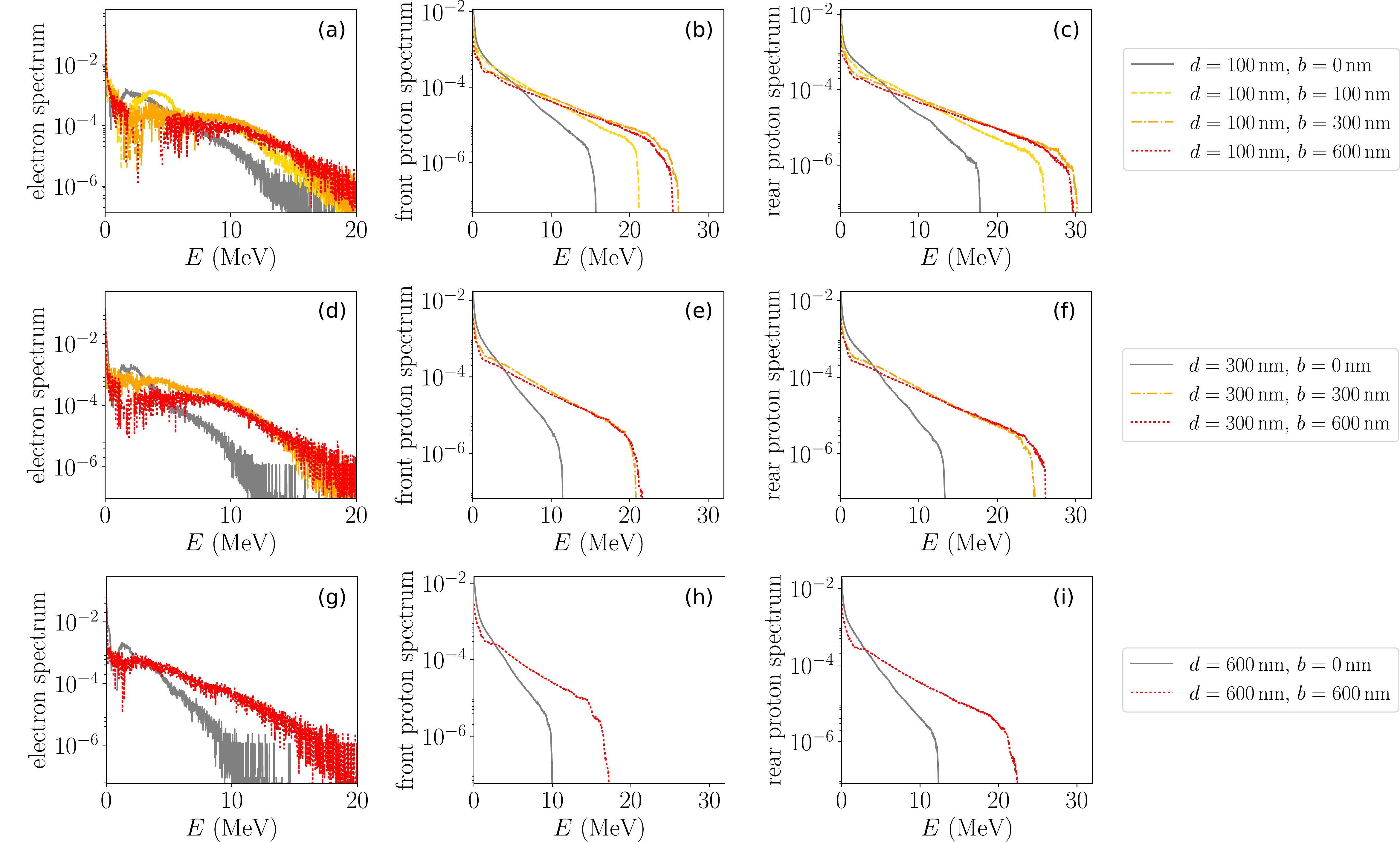}
	\caption{Particle density spectra of the rear-side electrons~(a,d,g), front-side protons~(b,e,h) and rear-side protons~(c,f,i) leaving the target. 
	The electron spectra are presented at $t=175$~fs, the front proton spectra at $t=315$~fs and the rear proton spectra at $t=455$~fs.
	The foil thickness is $d=100\,\mathrm{nm}$ in (a-c), $d=300\,\mathrm{nm}$ in (d-f), $d=600\,\mathrm{nm}$ in (g-i). The nanohole diameter is varied in the range $0\le b \le 600\,\rm nm$, as indicated in the legend of each row.
}
	\label{fig:spectra_bd}
\end{figure*}

We now perform a parametric scan where we vary the foil thickness $d$ and hole diameter $b$. Henceforth, we consider only the cases where the hole diameter is at least as large as the foil thickness, i.e., $b\geq d$.

Figure \ref{fig:spectra_bd}(a) presents the spectra of the rear-side electrons for a foil thickness from 100~nm to 600~nm and various hole diameters. One can see that the electrons from the flat foil~(gray line) reach lower energies than electrons from perforated foils, whatever the hole diameter. Especially above 5~MeV, perforated targets produce more high-energy electrons. The enhancement  is very similar for hole sizes from 100~nm to 600~nm. 

While we are mostly interested in the protons accelerated from the target backside, we also plot for completeness the energy spectra of the protons originating from the front side. Figures~\ref{fig:spectra_bd}(b,c) present the corresponding front~(b) and rear~(c) proton spectra. In both cases, the structuring significantly enhances the proton-cutoff energies. While a 100~nm hole size already enhances the proton-cutoff energy by about 40-50\%, the largest enhancements are reached for 300 to 600~nm wide holes.

When increasing the gold foil thickness, the trends remain the same~[see Figures~\ref{fig:spectra_bd}(d-i)]: The presence of the holes increases the amount of high-energy electrons and boosts the proton energy by about a factor of 2. The thinner the gold foil, the higher the proton energies. However, while this nanohole structuring can be used for all target thicknesses, the relative improvement  is slightly more pronounced for thicker foils.

The PIC simulation results suggest that as long as the
parameters are in the range leading to enhancement as described above,
they can be freely chosen as it is suitable from the point of view of
other experimental constraints, e.g.~ease of handling and fabrication.

Note that our simulations consider sharp-gradient targets, meaning that they do not address the possible influence of finite preplasmas caused by  laser prepulses. The generation of a significant preplasma would modify the picture: the holes might be filled with
electrons before the arrival of the main pulse, thus suppressing the benefit of the target structuring. Larger hole diameters may then be preferable in actual experimental conditions.

\begin{figure}
	\centering
	\includegraphics[width=0.99\columnwidth]{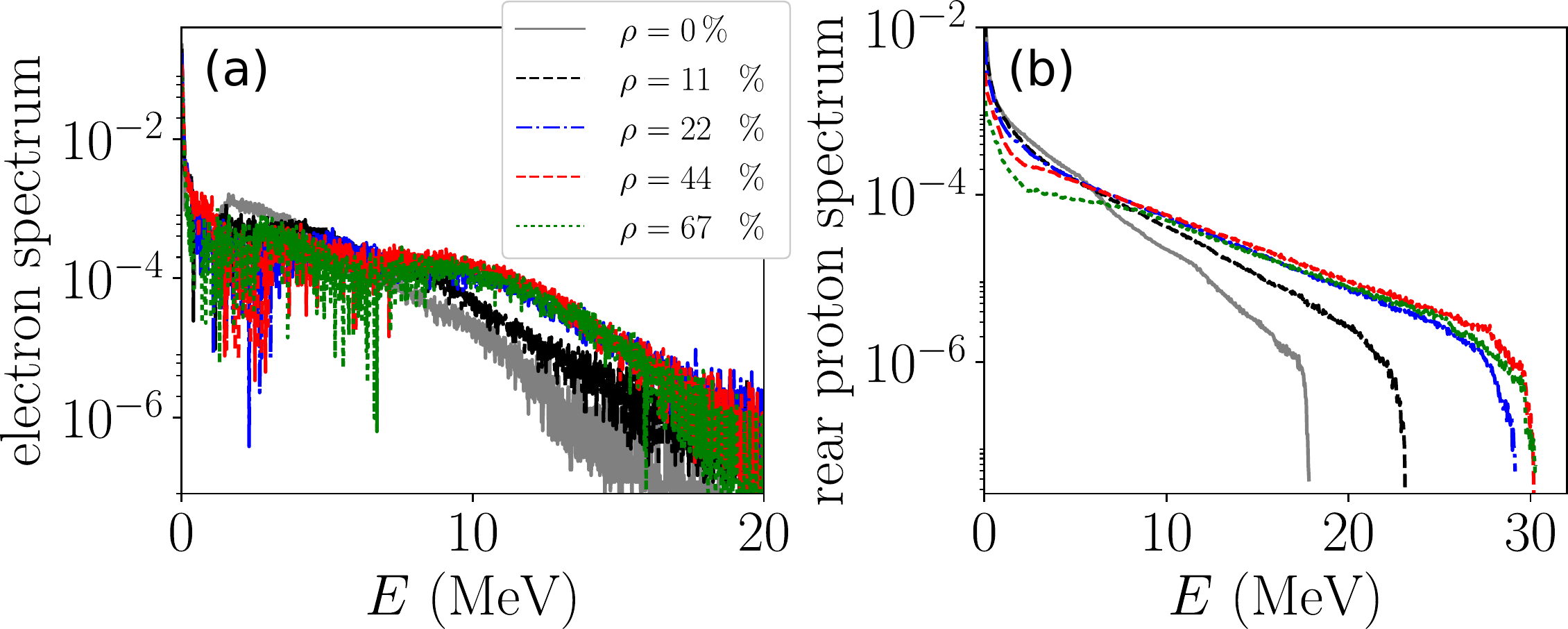}
	\caption{Rear-side electron and proton energy spectra under conditions similar to those of Fig.~\ref{fig:spectra_cones_vs_holes2}, but with different hole densities $\rho$ according to the legends in (a) and (b).}
	\label{fig:dens}
\end{figure}

The areal density of the holes can be expected to play a role in enhancing TNSA. Figure~\ref{fig:dens} displays the energy spectra of the rear-side electrons and protons obtained from $100\,\rm nm$-thick nanohole targets of nanohole density varying in the range $0\le\rho\le 67\%$. An optimum electron heating and proton acceleration is reached when almost half of the surface area is covered by the holes. However, already for a density of about $20\%$, a notable increase in the cutoff energy is reached. 

The optimum is likely formed by two counter-acting effects. On the one hand, when increasing the hole density, the mechanisms described in Sec.~\ref{sec:orig} further develop: more electrons can get an energy boost due to the increase in the interaction surface and recirculation. On the other hand, raising the hole density causes a decrease in the effective target volume, eventually resulting in higher laser transmission~(not shown here), and therefore reduced laser-target coupling. To mitigate the latter effect, one could use thicker foils instead of the considered 100~nm thin foils. However,  this is limited by the fact that the proton cutoff energy tends to be reduced when the nanohole-target gets thicker  -- in the same manner as for the standard flat foil target.

\section{\label{sec:sum}Conclusion}
\indent

In summary, by means of 2D PIC simulations, we investigate
laser-driven proton acceleration from non-periodic nanohole and
nanocone targets. We demonstrate a significant increase in the proton
cutoff energy in both types of structured targets compared to flat
foils. We show that the enhancement originates from a modification of
the interaction surface between the laser and the target, which allows
a higher number of electrons to be accelerated. Besides, the hot
electrons can interact with the laser pulse both longitudinally on the
front side of the target and transversely in the nanoholes, enabling
them to fully explore the phase-space, and be accelerated to higher
energies. A large parameter space in terms of hole diameter, foil
thickness and hole areal density yielding significant improvement of
the accelerated proton spectra is identified. We show that the
production of structured targets for improved ion acceleration can be
relaxed to non-periodic structures with a relatively low density of
structuring units.

\section*{Acknowledgements}
The authors acknowledge fruitful discussions with  L~Yi and the rest of the \textsc{PLIONA} team. This work was supported by the Knut and Alice Wallenberg Foundation, the European Research Council (ERC-2014-CoG grant 647121) and by the Swedish Research Council, Grant No. 2016-05012. The simulations were performed on resources provided by Grand Équipement National pour le Calcul Intensif (GENCI) (projects A0040507594, A0050506129) and at Chalmers Centre for Computational Science and Engineering (C3SE) provided by the Swedish National Infrastructure for Computing (grants SNIC2018-2-13 and SNIC2018-3-297).

\bibliographystyle{jpp}

\end{document}